\documentclass[letterpaper, 10 pt, conference]{ieeeconf}  

\IEEEoverridecommandlockouts                              
\usepackage{fullpage}
\usepackage{amssymb}
\usepackage{amsmath}
\usepackage{latexsym}
\usepackage{graphicx}
\usepackage{color}
\usepackage{url}

\usepackage{Shorthands}

\newcommand{\bsigma}{\bm{\sigma}}
\newcommand{\bvarepsilon}{\bm{\varepsilon}}
\usepackage{listings}
\usepackage{courier}
\usepackage{comment}
\usepackage[ruled]{algorithm2e}
\SetKwInput{KwInput}{Input}                
\SetKwInput{KwOutput}{Output}              
\makeatletter
\let\NAT@parse\undefined
\makeatother
\usepackage{hyperref}
\hypersetup{%
    pdfborder = {0 0 0}
}

\theoremstyle{definition}

\theoremstyle{definition}

\theoremstyle{remark}

\renewcommand{\hat}{\widehat}

\newcommand{\1}{{\mathbf 1}}

\newcommand{\er}{\textup{\textrm{er}}}

\usepackage{mathtools}

\newcommand{\charis}[1]{{\color{blue}{#1}}}
\newcommand{\vag}[1]{{\color{green}{#1}}}

\newcommand{\thickbar}[1]{\bm\bar{#1}}
\newcommand{\barf}{\ensuremath{\thickbar{f}}}

\title{\LARGE \bf Structural Risk Minimization for Learning Nonlinear Dynamics}

\author{Charis Stamouli, Evangelos Chatzipantazis, George J. Pappas
\thanks{The authors are with the Grasp Lab, University of Pennsylvania, Philadelphia, PA 19104, USA. Emails: \texttt{\{stamouli,vaghat,pappasg\}@seas.upenn.edu}.}
}

\begin{document}

\maketitle
\thispagestyle{empty}
\pagestyle{empty}

\begin{abstract}
Recent advances in learning or identification of nonlinear dynamics focus on learning a suitable model within a pre-specified model class.  However, a key difficulty that remains is the choice of the model class from which the dynamics  will be learned.  The fundamental challenge is trading the richness of the model class with the learnability within the model class.  Toward addressing the so-called model selection problem, we introduce a novel notion of Structural Risk Minimization (SRM) for learning nonlinear dynamics. Inspired by classical SRM for classification, we minimize a bound on the true prediction error over hierarchies of model classes. The class selected by our SRM scheme is shown to achieve a nearly optimal learning guarantee among all model classes contained in the hierarchy. Employing the proposed scheme along with computable model class complexity bounds, we derive explicit SRM schemes for learning nonlinear dynamics under hierarchies of: i) norm-constrained Reproducing Kernel Hilbert Spaces, and ii) norm-constrained Neural Network classes. We empirically show that even though too loose to be used as absolute estimates, our SRM bounds on the true prediction error are able to track its relative behavior across different model classes of the hierarchy. 
\end{abstract}

\section{Introduction}\label{introduction}

Modeling nonlinear dynamics to predict the future is of critical importance in a wide range of applications including e.g., fluid dynamics \cite{Anderson1995}, precision medicine \cite{Rajkomar2014}, and economics \cite{Day1994}. Conventionally, dynamical models are derived from first principles and data are used to estimate their parameters. However, physical systems are often governed by differential equations that are difficult to simulate numerically. Beyond that, mathematical models are not available for many real-world systems. Such systems have motivated the development of learning or identification of nonlinear dynamics, where system models are estimated entirely from data. 

The majority of works on identification of nonlinear dynamics \cite{Ghahramani1998,Langford2009,Khansari2011,Pillonetto2014,Svensson2017,Neumann2013,Raissi2018,Teng2019,Qin2021,Foster2020,Singh2021} focus on learning a suitable model within a pre-defined model class. The choice of the model class from which the dynamics will be learned is critical to the performance of the learned model. On one hand, a rich enough model class could contain the true dynamics model. On the other hand, learning in such a complex model class could be a very difficult task. Therefore, there is a fundamental trade-off in learning determined by the complexity of the selected model class. The model selection problem consists of choosing the model class that balances this trade-off.

To address the model selection problem, the authors in \cite{Brunton2016} propose a method for sparse identification of nonlinear dynamics. Their algorithm employs sparse regression and Pareto analysis to learn a parsimonious model from a combinatorial class of candidate models. In the extended version \cite{Mangan2017}, the most informative candidate models are ranked via appropriate information criteria. In both \cite{Brunton2016} and \cite{Mangan2017} the only assumption about the dynamics is that they can be sparsely represented. The authors in \cite{Champion2019} guarantee this assumption by learning an effective coordinate transformation to a suitable low-dimensional space. Exploiting the Koopman operator, recent works in \cite{Lusch2018,Kaiser2021} extend the scheme of \cite{Brunton2016} to classes of systems with continuous eigenvalue spectra.  

\begin{figure}[tbh]
   \centering
   \includegraphics[width=\linewidth]{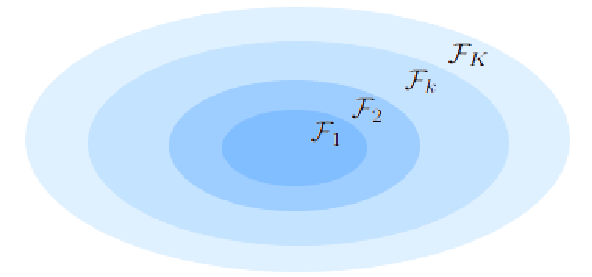}
   \caption{Illustration of a rich model class $\calF=\bigcup_{k=1}^K\calF_k$ hierarchized into $K$ nested subclasses $\calF_k$. For example, the classes $\calF_k$ could correspond to neural networks of increasing width.}
   \label{hierarchy_fig}
\end{figure}

In machine learning, a more foundational model selection approach is provided by Structural Risk Minimization (SRM) \cite{Lugosi1996,Koltchinskii2001,Shawe1998,Gyurik2023,Liu2020structural,Massucci2020,Meir1997}. SRM consists of: i) picking a hierarchy of model classes (see Figure~\ref{hierarchy_fig}), and ii) minimizing a class-dependent bound on the true prediction error over the entire hierarchy. In classical machine learning, SRM with strong learning guarantees has been proposed only for classification tasks, where data have binary labels. Particularly, in \cite{Lugosi1996,Koltchinskii2001} it is proved that the class corresponding to the respective
SRM classifier attains a nearly optimal learning trade-off. In system identification, SRM has been employed for learning switched autoregressive systems with external inputs, under an unknown number of modes \cite{Massucci2020}. In time series prediction, a general SRM scheme for scalar-valued stationary mixing processes has been developed \cite{Meir1997}.

In this paper, we introduce a novel notion of SRM for learning nonlinear dynamics. Our setting is different from the one in \cite{Meir1997} as it involves vector-valued non-stationary non-mixing stochastic processes. Given a hierarchy of model classes, our method estimates the true dynamics model by minimizing a bound on the true prediction error over the entire hierarchy. The bound is determined by the prediction error on the dataset and a penalty that captures the complexity of each class. The contributions of our work are the following:
\begin{enumerate}[label=\roman*)]
\item We propose the first SRM scheme for learning deterministic nonlinear dynamics with random initial condition, under arbitrary hierarchies of model classes.
\item We rigorously prove that the class selected by our SRM scheme achieves a nearly optimal learning guarantee among all model classes of the hierarchy.
\item We design explicit SRM schemes for our setting of nonlinear dynamics, under hierarchies of norm-constrained: a) Reproducing Kernel Hilbert Spaces (RKHSs), and b) Neural Network (NN) classes. This is achieved by combining our scheme with respective computable class complexity bounds.
\item We empirically show that even though too loose to be used as absolute estimates, our SRM bounds on the true prediction error are able to track its relative behavior across different model classes of the hierarchy. 
\end{enumerate}
All proofs can be found in Appendix~\ref{appendix_B}.

\section{Classical SRM in Machine Learning}\label{SRM_background}

Before introducing SRM for our setting of nonlinear dynamical systems (see Section~\ref{SRM_for_Learning_Nonlinear_Dynamics}), we first review classical SRM for the setting of classification \cite{Mohri,ShalevShwartz}.

The classification setting is characterized by an instance space $\calX\subseteq\setR^n$, the label space $\calY:=\{-1,+1\}$, an unknown probability distribution $\calD$ over $\calX\times\calY$, and an i.i.d. sample $S:=\{(x_i,y_i)\}_{i=1}^N\sim\calD^N$. Given the dataset $S$, the goal is to learn a classifier $h_S:\calX\to\calY$ that minimizes the $0-1$ true error (or risk), given by:
\begin{equation}\label{true_error}
    \er_\calD^{0-1}[h] = \Exp_{(x,y)\sim\calD}[\mathds{1}(h(x)\neq y)],
\end{equation}
for each $h:\calX\to\calY$, where $\mathds{1}(\cdot)$ denotes the indicator function. However, since the distribution $\calD$ is unknown, the true error \eqref{true_error} is not available to the learner. 

Empirical Risk Minimization (ERM) circumvents this issue by replacing the true error \eqref{true_error} with the corresponding training error:
\begin{equation}\label{training_error}
    \hat{\er}_S^{0-1}[h] := \frac{1}{N}\sum_{i=1}^N\mathds{1}(h(x_i)\neq y_i),
\end{equation}
which depends only on the dataset $S$. A key challenge that remains in ERM is model selection, that is, the problem of choosing the function class over which the training error \eqref{training_error} will be minimized. On one hand, a very rich function class is more likely to contain the ideal classifier $h_*(\cdot)$ with the minimum achievable risk. On the other hand, learning in such a complex function class could be a very difficult task. Consequently, there is a trade-off in learning determined by the complexity of the selected function class. To formalize this trade-off, let $\calH$ denote a class of functions from $\calX$ to $\calY$ and $h_S^{\ERM}(\cdot)$ the classifier returned by ERM on $\calH$. We can decompose the risk of $h_S^{\ERM}(\cdot)$ as $\er_\calD^{0-1}[h_S^{\ERM}]=\er_{\textup{app}}+\er_{\textup{est}}$,
where $\er_{\textup{app}}:=\min_{h\in\calH}\er_\calD^{0-1}[h]$ is the approximation error and $\er_{\textup{est}}:=\er_\calD^{0-1}[h_S^{\ERM}]-\er_{\textup{app}}$ is the estimation error.
The approximation error is the minimum risk achievable by a classifier in $\calH$ and depends on the richness of $\calH$. The estimation error measures how well the best classifier of $\calH$ is approximated by the ERM classifier $h_S^{\ERM}(\cdot)$ and depends on the dataset $S$. Notice that the choice of $\calH$ is subject to a fundamental trade-off between approximation and estimation errors: in particular, as the complexity of $\calH$ increases, the approximation error tends to decrease at the price of a larger estimation error. ERM overlooks this trade-off, which critically affects its practical performance.

SRM addresses the model selection problem, by replacing the true error \eqref{true_error} with a certain upper bound on it that allows regulating the approximation-estimation error trade-off.\vag{} This is because this bound is not determined only by the training error but also by some class complexity measure, such as Rademacher complexity. The Rademacher complexity captures the richness of a function class by measuring its ability to fit symmetric random noise. Herein, we will focus on the empirical Rademacher complexity $\hat{\calR}_S(\cdot)$ (see Appendix~\ref{appendix_A}), which depends on the sample $S$. In the theorem below, we present a Rademacher-based SRM scheme and a corresponding learning guarantee for classification. Note that the theorem below is a modified version of \cite[Theorem 4.2]{Mohri} for finite hierarchies of function classes.

\begin{theorem}[SRM for Classification]\label{th1}
Fix a failure probability $\delta\in(0,1)$. Let $\calH$ be a class of functions from $\calX$ to $\calY$ that can be written as $\calH=\bigcup_{k=1}^{K}\calH_k$ with $\calH_k\subset\calH_{k+1}$, $\forall k$. Then, for each $k$, with probability at least $1-\delta$ over the draw of $S\sim\calD^N$, $\calH_k$ satisfies:
\begin{equation}\label{ERM_UC_bound}
     \forall h\in\calH_k: \abs{\er_\calD^{0-1}[h]-\hat{\er}_S^{0-1}[h]}\leq\epsilon_k(S,\delta),
\end{equation}
where $\epsilon_k(S,\delta)=\hat{\calR}_S(\calH_k)+3\sqrt{\log(4/\delta)/(2N)}$. Moreover, if the SRM classifier is given by:
\begin{equation}\label{SRM_classifier}    
h_S^{\SRM}=\argmin_{1\leq k\leq K,h\in\calH_k}\left(\hat{\er}_S^{0-1}[h]+\hat{\calR}_S(\calH_k)\right),
\end{equation}
then, with probability at least $1-\delta$ over the draw of $S\sim\calD^N$, the following bound holds:
\begin{align}\label{SRM_guarantee}
    \er_{\calD}^{0-1}[h_S^{\SRM}]\leq\min\limits_{h\in\calH}\left(\er_{\calD}^{0-1}[h]+2\epsilon_{k(h)}\left(S,\frac{2\delta}{K+1}\right)\right),
\end{align}
where $k(h)=\min\{k\where h\in\calH_k\}$.
\end{theorem}

\noindent Notice in \eqref{SRM_classifier} that in contrast to ERM, SRM does not minimize just the training error, but the sum of the training error and the penalty $\hat{\calR}_S(\calH_k)$. Therefore, SRM trades some of its bias toward low training error with a bias toward classes $\calH_k$ for which $\hat{\calR}_S(\calH_{k})$ is small, for the sake of a smaller estimation error. The SRM guarantee \eqref{SRM_guarantee} ensures that the selected class achieves in fact a nearly optimal balance between approximation and estimation errors (see \cite{Mohri} for details).

\section{SRM for Learning Nonlinear Dynamics}\label{SRM_for_Learning_Nonlinear_Dynamics}

In the previous section, we summarized classical SRM for classification. We are now ready to introduce SRM for our setting of nonlinear dynamical systems. In Subsection~\ref{SRM_problem},  we formulate the problem of SRM for deterministic nonlinear dynamics with random initial condition. In Subsection~\ref{SRM_method}, we provide the first SRM scheme in the literature for this setting, along with a corresponding learning guarantee. 

 \subsection{SRM Problem for Nonlinear Dynamics}\label{SRM_problem}
 Consider a discrete-time nonlinear dynamical system of the form:
\begin{equation}\label{system}
    x_{t+1} = f_*(x_t),
\end{equation}
where $x_t\in\setR^{n}$ denotes the state at time $t$ and $f_*:\setR^n\to\setR^{n}$ the dynamics function, which is assumed to be unknown. The initial state $x_0$ is modeled as a random variable with an unknown probability distribution $\calD$. Let $T\geq1$ be the time horizon over which trajectories generated by system \eqref{system} are considered (e.g., $T$ could be the system operation time). Moreover, let $x_{0:T}(\xi)=\begin{bmatrix}x_0(\xi),\ldots,x_T(\xi)\end{bmatrix}^\intercal$, where $x_t(\xi)$ denotes the state of system \eqref{system} at time $t$, for some initial condition $\xi\sim\calD$.

Suppose we are given a set $S:=\{x_{0:T}(\xi_i)\}_{i=1}^N$ of $N$ trajectories of system \eqref{system} corresponding to an i.i.d. sample of initial conditions $S_0:=\{\xi_i\}_{i=1}^N\sim\calD^N$. Note the implicit dependence of $S$ on $S_0$. Given the dataset $S$, the objective is to learn a predictor $f_S:\setR^n\to\setR^n$ such that for each $\xi\sim\calD$, we have $f_S(x_t(\xi))\approx f_*(x_t(\xi))$, for all $t=0,\ldots,T-1$, that is, $f_S(\cdot)$ should be able to approximate the next state at each time step $t$. Formally, given the dataset $S$ and a loss function $\ell:\setR^n\to\setR_+$, we seek a predictor $f_S:\setR^n\to\setR^n$ that minimizes the corresponding true error (or risk), given by:
\begin{equation}\label{true_error1}
    \er_{\calD}[f] = \Exp_{\xi\sim\calD}\left[\frac{1}{T}\sum_{t=0}^{T-1}\ell\left(f(x_t(\xi))-x_{t+1}(\xi)\right)\right],
\end{equation}
for each $f:\setR^n\to\setR^n$.
\begin{assumption}\label{ass_loss}
The loss function $\ell(\cdot)$ is locally Lipschitz continuous and satisfies $\ell(0)=0$.   \end{assumption}
\noindent The above assumption is satisfied by all standard regression loss functions (e.g., $L_p$ loss for each $p\geq1$, Huber loss). Notice that the true error \eqref{true_error1} cannot be minimized directly as it depends on the unknown distribution $\calD$. As in the case of classification (see Section~\ref{SRM_background}), we can tackle this challenge by employing ERM, which replaces the true error \eqref{true_error1} with the corresponding training error:
\begin{equation}\label{training_error1}
   \hat{\er}_S[f]:=\frac{1}{NT}\sum_{i=1}^N\sum_{t=0}^{T-1}\ell\left(f(x_t(\xi_i))-x_{t+1}(\xi_{i})\right).
\end{equation}

\begin{assumption}\label{ass_sys}
There exists a known bound $B>0$ such that $\norm{x_{t}(\xi)}_2\leq B$, for all $t=0,\ldots,T$ and $\xi\sim\calD$.
\end{assumption}

\begin{remark}\label{rem_sys}
The above assumption implies that the predicted states should lie in the Euclidean ball of radius $B$. To guarantee that, we employ the method of clipping, which is commonly used in machine learning to obtain bounded predictors by searching over unbounded model classes \cite{Mohri}. In particular, for each predictor $f:\setR^n\to\setR^n$, we define its clipped version $\thickbar{f}(\cdot)$ as:
\begin{equation}\label{normalized_predictor}
    \thickbar{f}(x) = \left\{
            \begin{array}{lcr}
            \frac{B}{\norm{f(x)}_2}\cdot f(x), \text{ if } \norm{f(x)}_2> B\\
               \hspace{1.3cm}f(x), \text{ else}
               
            \end{array}
            \right., 
\end{equation}
for all $x\in\setR^n$. Note that in practice the constant $B$ could be a rough bound that we know from physics or have obtained from observations.
\end{remark}

In classical nonlinear system identification, ERM is employed on some function class $\calF$ which is assumed to contain the true dynamics model \cite{Ziemann2022}. This assumption, though, is typically unrealistic in practice, unless the selected function class $\calF$ is very rich. However, learning in a highly complex function class could be a difficult or even intractable task. Hence, there is a trade-off in learning regulated via the complexity of $\calF$. In Section~\ref{SRM_background}, we formalized this trade-off by introducing the notions of the approximation and estimation errors. Particularly, recall that as the complexity of the class increases, the approximation error decreases at the price of a larger estimation error. But then, how should $\calF$ be chosen? 

In this paper, we address the above question by
formulating SRM for learning nonlinear dynamics. More specifically, our goal is to design a learning algorithm that estimates $f_*(\cdot)$ with a predictor that guarantees a nearly optimal approximation-estimation error trade-off.

\begin{problem}\label{problem1}
Let $\calF$ be a class of functions from $\setR^n$ to $\setR^n$ that can be written as $\calF=\bigcup_{k=1}^{K}\calF_k$ with $\calF_k\subset\calF_{k+1}$, $\forall k$. Moreover,
let $L>0$ be the Lipschitz constant of $\ell(\cdot)$ on the set $Z:=\{x\in\setR^n\where\norm{x}_2\leq2B\}$. Given a failure probability $\delta\in(0,1)$, for each $k$:
\begin{enumerate}[label=\roman*)]
\item Derive a bound $\epsilon_k(S,\delta)$ such that with probability at least $1-\delta$ over the draw of $S_0\sim\calD^N$:
\begin{equation}\label{ERM_UC_bound1}
     \forall f\in\calF_k: \abs{\er_\calD[\thickbar{f}]-\hat{\er}_S[\thickbar{f}]}\leq\epsilon_k(S,\delta).
\end{equation}
\item Design a penalty $r_k(S)$ such that with probability at least $1-\delta$ over the draw of $S_0\sim\calD^N$, the SRM predictor:
\begin{equation}\label{SRM_predictor}    
f_S^{\SRM}:=\argmin_{1\leq k\leq K,f\in\calF_k}\left(\hat{\er}_S[\thickbar{f}]+r_k(S)\right)
\end{equation}
satisfies:
\begin{align}\label{SRM_guarantee1}
    \er_{\calD}[\thickbar{f}_S^{\SRM}]\leq\min\limits_{f\in\calF}\Big(\er_{\calD}[\thickbar{f}]+2\epsilon_{k(f)}\Big(S,\frac{2\delta}{K+1}\Big)\Big),
\end{align}
where $k(f)=\min\{k\where f\in\calF_k\}$.
\end{enumerate}
The bounds $\epsilon_k(S,\delta)$ and the penalties $r_k(S)$ may also depend on the parameters $n$, $N$, $T$, $B$, $L$. 
\end{problem}

\begin{remark}\label{rem_nested}
For clarity of presentation, in Problem~\ref{problem1} we assumed that the classes $\calF_k$ are nested, as dictated by the condition $\calF_k\subset\calF_{k+1}$, $\forall k$. However, note that our results directly extend to arbitrary decompositions of $\calF$. 
\end{remark}

\subsection{SRM Method for Nonlinear Dynamics}\label{SRM_method}
In this subsection, we introduce the first SRM scheme for nonlinear dynamics under the setting described in Subsection~\ref{SRM_problem}. Similarly to classical SRM (see Theorem~\ref{th1}), for each $k$, we provide a Rademacher-based penalty $r_k(S)$ for the SRM predictor \eqref{SRM_predictor}. The main differences between our setting and that of classical SRM are the following:
\begin{enumerate}[label=\roman*)]
\item In our setting, the function that we wish to estimate is \textit{vector-valued}, whereas in classical SRM, it is \textit{scalar-valued}. Particularly, note that $f_*(\cdot)$ takes values in $\setR^n$, whereas $h_*(\cdot)$ takes values in $\{-1,+1\}$.
\item Our dataset consists of $N$ \textit{i.i.d. sequences of dependent datapoints} (trajectories). In classical SRM, the dataset consists of $N$ \textit{i.i.d. datapoints}.
\item The risk \eqref{true_error1} employed in our setting is defined as the expected \textit{average $\ell$-loss over trajectories of length $T$}, whereas the risk \eqref{true_error} employed in classical SRM is simply defined as the expected \textit{$0-1$ loss on a single datapoint}.
\end{enumerate}
Consequently, our framework provides an extension of classical SRM to vector-valued predictors, i.i.d. sequences of data, and a large family of losses, suitably designed for the setting of nonlinear dynamics. Our setting is also different from the one in \cite{Meir1997}, where \textit{scalar-valued} time series are considered under extra \textit{stationarity} and \textit{mixing} assumptions. In the theorem below, we present our Rademacher-based SRM scheme along with a corresponding learning guarantee. 

\begin{theorem}[SRM for Nonlinear Dynamics]\label{th2}
Fix a failure probability $\delta\in(0,1)$. Let $\calF$ be a class of functions from $\setR^n$ to $\setR^n$ that can be written as $\calF=\bigcup_{k=1}^{K}\calF_k$ with $\calF_k\subset\calF_{k+1}$, $\forall k$. Moreover, let the SRM predictor $f_S^{\SRM}(\cdot)$ be as in \eqref{SRM_predictor} with:
\begin{equation}\label{penalty}
    r_k(S)=\frac{2\sqrt{2}L}{T}\sum_{t=0}^{T-1}\sum_{j=1}^n\hat{\calR}_{S_t}(\thickbar{\calF}_{k,j}),
\end{equation}
where $\thickbar{\calF}_{k,j}=\left\{j\text{-th component of }\barf(\cdot)\where f\in\calF_k\right\}$ and $S_t=\{x_t(\xi_i)\}_{i=1}^N$. Then, setting $\epsilon_k(S,\delta)=r_k(S)+6LB\sqrt{\log(4/\delta)/(2N)}$, each of \eqref{ERM_UC_bound1} and \eqref{SRM_guarantee1} holds with probability at least $1-\delta$ over the draw of $S_0\sim\calD^N$.
\end{theorem}

\noindent Note that the definition of the empirical Rademacher complexity applies to bounded classes of scalar-valued functions and i.i.d. datasets, which is not the case in our setting. To tackle this challenge, in the theorem above we employed the clipped component classes $\thickbar{\calF}_{k,j}$ and the i.i.d. sub-datasets $S_t$. Employing a suitable contraction inequality from \cite{Cortes2016}, we obtained the penalty \eqref{penalty}, which depends on the empirical Rademacher complexities $\hat{\calR}_{S_t}(\thickbar{\calF}_{k,j})$. As expected, the penalty \eqref{penalty} is more complicated than the one of classical SRM (see \eqref{SRM_classifier}). 

In the following remark, we show that the SRM predictor of Theorem~\ref{th2} guarantees a nearly optimal balance between approximation and estimation errors.

\begin{remark}[SRM Guarantee for Nonlinear Dynamics]\label{remark_SRM_guarantee}
Suppose the true dynamics function $f_*(\cdot)$ is contained in $\calF=\bigcup_{k=1}^K\calF_k$. Let $k(f_*):=\min\left\{k:f_*\in\calF_k\right\}$ be the index of the class that optimizes the approximation-estimation error trade-off. Moreover, let $f_{S}^{\ERM}(\cdot)$ be the ERM predictor in $\calF_{k(f_*)}$. Then, from 
\eqref{ERM_UC_bound1} and \cite[Proposition 4.1]{Mohri} we obtain:
\begin{equation}\label{ERM_guarantee}
\er_\calD[\barf_{S}^{\ERM}]\leq\er_\calD[\barf_*]+2\,\epsilon_{k(f_*)}(S,\delta),
\end{equation}
whereas \eqref{SRM_guarantee1} yields:
\begin{equation}\label{SRM_star_guarantee}
\er_\calD[\barf_S^{\SRM}]\leq\er_{\calD}[\barf_*]+2\,\epsilon_{k(f_*)}(S,(2\delta)/(K+1)). 
\end{equation}
From \eqref{ERM_guarantee} and \eqref{SRM_star_guarantee} we conclude that the bound on the risk of the clipped SRM predictor $\barf_S^{\SRM}(\cdot)$ is similar to that of the clipped ERM predictor $\barf_{S}^{\ERM}(\cdot)$: the only difference is a factor of $2/(K+1)$ on the failure probability $\delta$. Hence, modulo that factor, the SRM guarantee \eqref{SRM_guarantee1} is as favorable as the one we would have obtained, had an oracle informed us a priori of the optimal index $k(f_*)$.
\end{remark}

Beyond the above theoretical guarantee, SRM provides us with a model selection framework that applies to classes of arbitrary parametric model structures. The structural nature of SRM will become more clear in the next section, where explicit SRM schemes are provided for hierarchies of RKHS and NN classes.

\section{Explicit SRM for RKHS and NN Classes}\label{explicit_SRM}
The general SRM scheme we proposed in Section~\ref{SRM_for_Learning_Nonlinear_Dynamics} for our setting of nonlinear dynamics consists of: i) picking a rich class $\calF=\bigcup_{k=1}^{K}\calF_k$ with $\calF_k\subset\calF_{k+1}$, $\forall k$, and ii) minimizing the sum of the training error \eqref{training_error1} and the penalty \eqref{penalty}, as dictated by \eqref{SRM_predictor}. However, computing the penalty \eqref{penalty} is typically impossible in practice, as an exact expression of the empirical Rademacher complexities $\hat{\calR}_{S_t}(\thickbar{\calF}_{k,j})$ is generally unavailable. To tackle this issue, we could replace $\hat{\calR}_{S_t}(\thickbar{\calF}_{k,j})$'s with respective \textit{computable} upper bounds. In this section, we employ such bounds and derive explicit SRM schemes for two large families of nonlinear functions: i) norm-constrained Reproducing Kernel Hilbert Spaces (RKHSs), and ii) norm-constrained Neural Network (NN) classes.

Before presenting our explicit SRM schemes, let us first introduce some notation. Consider the setting of Problem~\ref{problem1} and for each $k$, let $\barf_k(\cdot)$ be the clipped version of $f_k:=\argmin_{f\in\calF_k}\hat{\er}_S[\barf]$. Herein, we refer to $\barf_k(\cdot)$ as the \textit{$k$-class predictor} of $\calF$. For each $k$, let us also define the SRM error of $\barf_k(\cdot)$ as $\er_{\SRM}[\barf_k]=\hat{\er}_S[\barf_k]+r_k(S)$. Notice that the clipped SRM predictor is the $k$-class predictor with the minimum SRM error.

\subsection{RKHS-based SRM for Nonlinear Dynamics}\label{RKHS-based_SRM_for_Nonlinear_Dynamics}
In this subsection, we introduce the norm-constrained RKHS \cite{Mohri} class employed in our setting as well as a corresponding SRM scheme.

Let $\kappa_\theta:\setR^n\times\setR^n\to\setR$ be a continuous positive definite symmetric kernel \cite[Definition 6.3]{Mohri} parametrized by a real vector $\theta$. Examples of commonly used kernels \cite{Bishop} include the polynomial kernel $\kappa_{c,q}(x,x'):=\left(x^\intercal x'+c\right)^q$, where $c\in\setR_+$ and $q\geq1$, and the Gaussian kernel $\kappa_{\gamma}(x,x'):=\exp(-\gamma\norm{x-x'}_2^2)$, where $\gamma>0$. Let $\setH$ denote the RKHS associated to $\kappa_{\theta}$ and $\norm{\cdot}_{\setH}$ its corresponding norm \cite{Mohri}. For any $\calB>0$, we define the \textit{$\calB$-constrained RKHS with kernel $\kappa_{\theta}$} as $\calF_{\RKHS}(\kappa_{\theta},\calB)=\big\{f:=\begin{bmatrix}f_1,\ldots,f_n\end{bmatrix}^\intercal\where f_j\in\setH,\norm{f_j}_{\setH}\leq \calB,\forall j\big\}$. 

Consider a hierarchy of classes $\calF_k=\calF_{\RKHS}(\kappa_{\theta_k},\calB_k)$ of the above form. In that case, the SRM scheme of Theorem~\ref{th2} cannot be applied directly, since an exact expression of $\hat{\calR}_{S_t}(\thickbar{\calF}_{k,j})$'s is not available. However, we can upper-bound $\hat{\calR}_{S_t}(\thickbar{\calF}_{k,j})$'s using standard computable bounds from \cite{Mohri}. Employing these bounds, we obtain the RKHS-based SRM scheme of the theorem below (see Appendix~\ref{appendix_C} for implementation details).

\begin{theorem}[RKHS-based SRM for Nonlinear Dynamics]\label{th3}
Fix a failure probability $\delta\in(0,1)$. Let $\calF=\bigcup_{k=1}^K\calF_k$ with $\calF_k=\calF_{\RKHS}(\kappa_{\theta_k},\calB_k)$, $\forall k$, where the parameters $\kappa_{\theta_k}$, $\calB_k$ are such that $\calF_k\subset\calF_{k+1}$, $\forall k$. Let $\setH_k$ be the RKHS associated to $\kappa_{\theta_k}$ and $\{M_1,\ldots,M_Q\}$ an increasing sequence of positive values with $M_Q=\max_{k}\calB_{k}$. For each $f\in\calF$, we define $q(f)=\min\left\{q\,\big|\,\max_j\norm{f_j}_{\setH_{k(f)}}\leq M_q\right\}$, where $k(f)=\min\{k\where f\in\calF_k\}$. Suppose the SRM predictor $f_S^{\SRM}(\cdot)$ is given by \eqref{SRM_predictor} with: 
\begin{equation}\label{RKHS_penalty}  
    r_k(S) = \frac{2\sqrt{2}LnM_{q(f_k)}}{TN}\sum\limits_{t=0}^{T-1}\sqrt{\sum\limits_{i=1}^N\kappa_{\theta_k}\left(x_t(\xi_i),x_t(\xi_i)\right)}.
\end{equation}
Then, if $\epsilon_k(S,\delta)=\tilde{r}_k(S)+6LB\sqrt{\log(4Q/\delta)/(2N)}$, where $\tilde{r}_k(S)$ is defined as $r_k(S)$ but with $M_{q(f_k)}$ replaced by $M_{\max\{q(f_k),q(f)\}}$, each of \eqref{ERM_UC_bound1} and \eqref{SRM_guarantee1} holds with probability at least $1-\delta$ over the draw of $S_0\sim\calD^N$.
\end{theorem}

\noindent Note that the penalty \eqref{RKHS_penalty} does not rely on the worst-case constraint constant $\calB_k$ of the class $\calF_k$, but on the discretized value $M_{q(f_k)}$ of the \textit{learned} model $f_k(\cdot)$, which is selected from a pre-defined set $\{M_1,\ldots,M_Q\}$.

\subsection{NN-based SRM for Nonlinear Dynamics}\label{NN-based_SRM_for_Nonlinear_Dynamics}
In this subsection, we present the class of neural networks \cite{ShalevShwartz} employed in our setting, along with a corresponding SRM scheme.

Let $g:\setR\to\setR$ be a $1$-Lipschitz function that satisfies $g(az)=ag(z)$, for all $a\in\setR_+$ and $z\in\setR$ (e.g., consider the ReLU function $g(z):=\max\{0,z\}$, $\forall z$). Somewhat abusing notation, we will also allow $g(\cdot)$ to be applied element-wise (i.e., applied over each coordinate of its input). Let us also define $\tilde{g}=\begin{bmatrix}g&\boldsymbol{1}\end{bmatrix}^\intercal$, where $\boldsymbol{1}(\cdot)$ denotes the constant function equal to $1$. Consider a neural network of the form:
\begin{equation}\label{NN_model}
f(x)=W_{f}^{(D_f)}\tilde{g}\left(W_{f}^{(D_f-1)}\tilde{g}\left(\cdots\tilde{g}\left(W_{f}^{(1)}\begin{bmatrix}
        x\\1
    \end{bmatrix}\right)\right)\right),
\end{equation}
where $x\in\setR^n$ is the input, $D_f\geq1$ is the depth, and $W_f^{(1)},\ldots,W_f^{(D_f)}$ are the weight matrices of the neural network. Suppose we have $W_{f}^{(1)}\in\setR^{(H_f+1)\times(n+1)}$, $W_{f}^{(D_f)}\in\setR^{n\times(H_f+1)}$, and $W_{f}^{(d)}\in\setR^{(H_f+1)\times(H_f+1)}$, for all $d=2,\ldots,D_f-1$, where $H_f\geq1$ is the width of the neural network. Notice
the increase of the input dimension $n$ and the width $H_f$ in the weight matrices $W_f^{(d)}$, so that biases are incorporated in them. For any $D\geq1$, $H\geq1$ and $\calB>0$, we define the \textit{class of $\calB$-constrained neural networks of depth $D$ and width $H$} as:
\begin{align}\label{NN_class}
 &\calF_{\NN}(D,H,\calB)=\Big\{x\mapsto f(x)\,\Big|\,f(\cdot) \text{ satisfies \eqref{NN_model},}\,\forall x,\nonumber\\&  D_f=D, H_f=H,\norm{W_f^{(d)}}_{\F}\leq\calB,\forall d\Big\}.  
\end{align}
In the case of NN classes $\calF_k=\calF_{\NN}(D_k,H_k,\calB_k)$ of the form \eqref{NN_class},  an exact expression of $\hat{\calR}_{S_t}(\thickbar{\calF}_{k,j})$'s is not available. Therefore, the SRM scheme of Theorem~\ref{th2} cannot be applied directly. Nevertheless, we can upper-bound $\hat{\calR}_{S_t}(\thickbar{\calF}_{k,j})$'s using standard computable bounds from \cite{Golowich2019}. Employing theses bounds, we derive the NN-based SRM scheme presented in the following theorem.

\begin{theorem}[NN-based SRM for Nonlinear Dynamics]\label{th4}
Fix a failure probability $\delta\in(0,1)$. Let $\calF=\bigcup_{k=1}^K\calF_k$ with $\calF_k=\calF_{\NN}(D_k,H_k,\calB_k)$, $\forall k$, where the parameters $D_k$, $H_k$, $\calB_k$ are such that $\calF_k\subset\calF_{k+1}$, $\forall k$. Moreover, let $\{M_1,\ldots,M_Q\}$ be an increasing sequence of positive values with $M_Q=\max_{k}\calB_{k}$. For each $f\in\calF$, we define $q(f)=\min\left\{q\,\big|\,\max_d\norm{W_f^{(d)}}_{\F}\leq M_q\right\}$. Suppose the SRM predictor $f_S^{\SRM}(\cdot)$ is given by \eqref{SRM_predictor} with: 
\begin{align}\label{NN_penalty}
  r_k(S)=\frac{LnM_{q(f_k)}^{D_k}a(D_k)}{TN}\sum_{t=0}^{T-1}\sqrt{\sum_{i=1}^N(\norm{x_t(\xi_i))}_2^2+1)},
\end{align}
where $a(D_k)=2\sqrt{2}\,(\sqrt{2\log(2)D_k}+1)$. Then, if $\epsilon_k(S,\delta)=\tilde{r}_k(S)+6LB\sqrt{\log(4Q/\delta)/(2N)}$, where $\tilde{r}_k(S)$ is defined as $r_k(S)$ but with $M_{q(f_k)}$ replaced by $M_{\max\{q(f_k),q(f)\}}$, each of \eqref{ERM_UC_bound1} and \eqref{SRM_guarantee1} holds with probability at least $1-\delta$ over the draw of $S_0\sim\calD^N$.
\end{theorem}
\noindent Note that the penalty \eqref{NN_penalty}
does not rely on the worst-case constraint constant $\calB_k$ of the class $\calF_k$, but on the discretized value $M_{q(f_k)}$ of the \textit{learned} model $f_k(\cdot)$, which is selected from a pre-defined set $\{M_1,\ldots,M_Q\}$.

\begin{remark}\label{rem_discret}
Unlike the general penalty \eqref{penalty}, the RKHS penalty \eqref{RKHS_penalty} and the NN penalty \eqref{NN_penalty} are both computable, thus providing us with the first explicit SRM schemes for learning of nonlinear dynamics. Note the implicit dependence of \eqref{RKHS_penalty} on $\kappa_{\theta_k}$ and of \eqref{NN_penalty} on $H_k$ via the discretized value $M_{q(f_k)}$ of the respective learned model $f_k(\cdot)$. Similarly to our general SRM scheme, our RKHS- and NN-based SRM schemes guarantee a nearly optimal trade-off between approximation and estimation errors (see Remark~\ref{remark_SRM_guarantee}). Due to discretization through the pre-specified set $\{M_1,\ldots,M_Q\}$, the corresponding error bounds $\epsilon_k(S,\delta)$ involve an extra factor of $1/Q$ on the failure probability $\delta$. However, the failure probability does not affect the penalties, which in fact become tighter as $Q$ increases. Hence, one could choose the discretized values $M_1,\ldots,M_{Q-1}$ by splitting $\begin{bmatrix}0,M_Q\end{bmatrix}$ into sub-intervals of equal length, depending on the desired numerical accuracy.
\end{remark}

\section{Experiments}\label{Experiments}
In this section, we illustrate the efficacy of the proposed SRM schemes using two classical nonlinear systems. Specifically, in Subsection~\ref{double_pendulum}, we estimate the double pendulum using our RKHS-based SRM scheme, whereas in Subsection~\ref{triple_pendulum}, we estimate the triple pendulum using our NN-based SRM scheme. Both pendula consist of identical compound pendula of length $l=1$ and mass $m=0.5$. Furthermore, in each case, we obtain the corresponding discretization set $\{M_1,\ldots,M_Q\}$ by splitting $\begin{bmatrix}0,M_Q\end{bmatrix}$ into sub-intervals of length $0.01$.

\subsection{Double Pendulum}\label{double_pendulum}
Assume the double pendulum's motion is modeled via Hamilton's equations \cite{Stickler2016}, with each component of the initial state independently uniformly distributed in $\begin{bmatrix}-0.5,0.5\end{bmatrix}$. The system is $4$-dimensional with the first two states representing the angular position and the last two states the momentum \cite{Stickler2016}. The bound $B$ on the system state norm was roughly computed as $B=2.2$ (see Remark~\ref{rem_sys}). Suppose we are given a dataset $S$ of $N=140$ trajectories of length $T=5$. We generated these trajectories by sampling the original (continuous-time) trajectories with period $T_s=0.05s$. Employing the dataset $S$ and the loss function $\norm{\cdot}_2$, our goal is to estimate the discretized model $f_*(\cdot)$ using our RKHS-based SRM scheme.
In particular, consider Gaussian kernels with parameters $\gamma_k=10^{-5},10^{-4},10^{-3},10^{-2},10^{-1},1,5,10$ and define the class $\calF=\bigcup_{k=1}^8\calF_k$ with $\calF_k=\calF_{\RKHS}(\kappa_{\gamma_k},20)$, $\forall k$. Notice that the complexity of $\calF_k$ increases with $k$, despite the fact that the classes $\calF_k$ are not nested (see Remark~\ref{rem_nested}). This is because smaller values of $\gamma$ produce Gaussian kernels with smoother decision boundaries.

\begin{figure}[tbh]
   \centering
   \includegraphics[width=\linewidth]{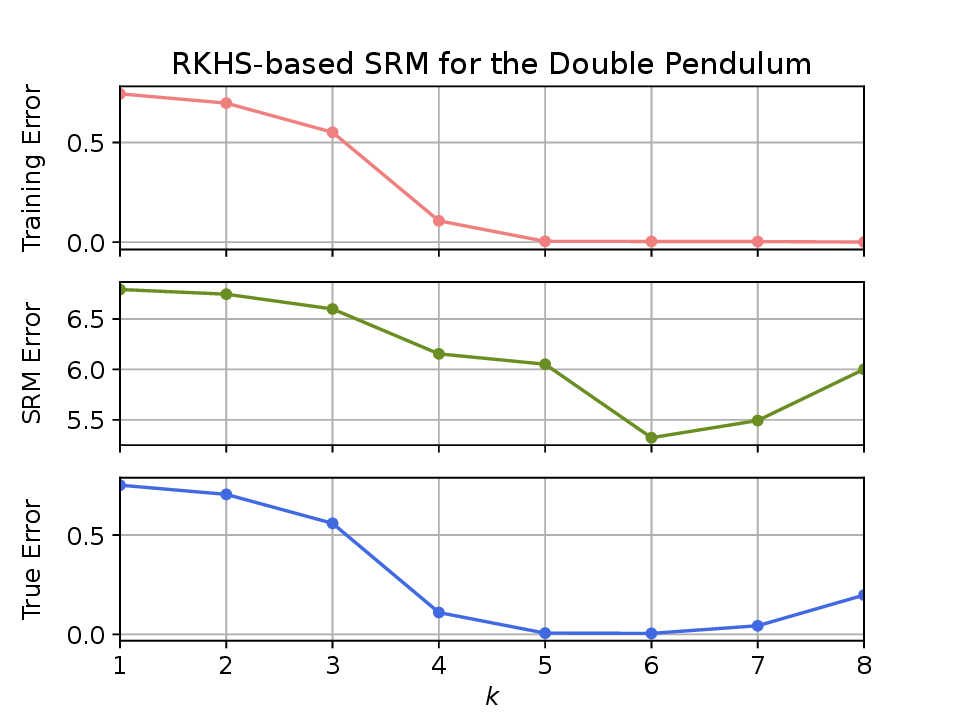}
   \caption{The training error (top), the SRM error (middle), and the true error (bottom) of the $k$-class predictors $\barf_k(\cdot)$ corresponding to the RKHS classes $\calF_{\RKHS}(\kappa_{\gamma_k},20)$. The training error fails to track the increase of the true error for $k>6$, whereas the SRM error achieves tracking the relative behavior of the true error across all $k$'s. }
   \label{RKHS_fig}
\end{figure}

Figure~\ref{RKHS_fig} shows the training, SRM and true errors of the $k$-class predictors $\barf_k(\cdot)$ learned from the RKHS classes $\calF_k$. We estimated the true error using $10^4$ test trajectories. We observe that the true error decreases for $k<6$ and increases for $k>6$. This was expected since smaller $\gamma$'s lead to simpler classes that may not capture the complexity of $f_*(\cdot)$, whereas larger $\gamma$'s lead to more complex classes where learning is more difficult. We can see that the training error keeps decreasing over $k$, thus failing to follow the true error for $k>6$. In contrast, the SRM error tracks the relative behavior of the true error for all $k$. Particularly, note that the minimum SRM and true error are both attained at $\gamma_6=1$, which yields a risk of $\er_{\calD}[\barf_S^{\SRM}]\approx0.006$. 

\subsection{Triple Pendulum}\label{triple_pendulum}
Assume the triple pendulum's motion is modeled via Kane's equations \cite{Gede2013}. The system is $6$-dimensional with the first three states describing the angular position and the last three states the angular velocity. The initial state is drawn from a zero-mean Gaussian truncated at $B_{\xi}:=\{x\in\setR^6\where\abs{x_i}\leq\pi/4,i=1,2,3,\abs{x_i}\leq0.5,i=4,5,6\}$. The bound on the system state norm was roughly computed as $B=4$ (see Remark~\ref{rem_sys}). Suppose we are given a dataset $S$ of $N=150$ trajectories of length $T=5$. We generated these trajectories by sampling the original (continuous-time) trajectories with period $T_s=0.05$. Employing the dataset $S$ and the loss function $\norm{\cdot}_2$, we wish to estimate the discretized model $f_*(\cdot)$ using our NN-based SRM scheme. Specifically, consider neural networks with ReLU activation, constant depth $D=2$, and increasing width $H_k=25,35,\ldots,95$, and define the class $\calF=\bigcup_{k=1}^8\calF_k$ with $\calF_k=\calF_{\NN}(2,H_k,12)$, $\forall k$.

\begin{figure}[tbh]
   \centering
   \includegraphics[width=\linewidth]{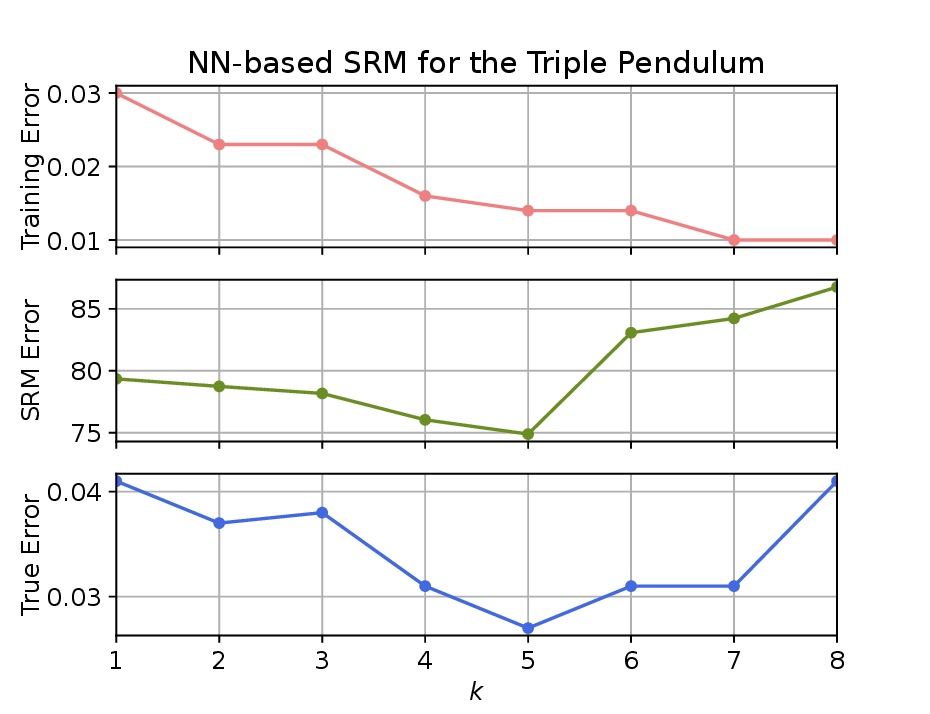}
   \caption{The training error (top), the SRM error (middle), and the true error (bottom) of the $k$-class predictors $\barf_k(\cdot)$ corresponding to the NN classes $\calF_{\NN}(2,H_k,12)$. The training error fails to track the increase of the true error for $k>5$. In contrast, the SRM error achieves tracking the relative behavior of the true error across all $k$'s (except for $k=3$ where a negligible increase of the true error is observed).}
   \label{NN_fig}
   \end{figure}

Figure~\ref{NN_fig} shows the training, SRM and true errors of the $k$-class predictors $\barf_k(\cdot)$ learned from the NN classes $\calF_k$. We implemented the norm-constrained NN classes as in \cite{Gouk2021} and estimated the true error using a dataset of $10^4$ test trajectories. Notice that the true error decreases for $k<5$ and increases for $k>5$ (we will ignore the negligible increase at $k=3$). This is because neural networks with smaller width may not capture the complexity of $f_*(\cdot)$, whereas neural networks with larger width are more difficult to learn. We observe that the training error decreases for the entire range of NN widths, thus failing to follow the true error for $k>5$. Unlike the training error, the SRM error achieves tracking the relative behavior of the true error for all $k$. Particularly, note that the minimum SRM and true error are both attained at $H_5=65$, which yields a risk of $\er_{\calD}[\barf_S^{\SRM}]\approx0.027$.
\vspace{0.15cm}

\noindent\textbf{Discussion:} In both experiments we observe that the values of the SRM error are of larger order than the values of the training and true error. Despite being too loose to be used as absolute estimates of the true error, our SRM bounds remarkably succeed in tracking its relative behavior. Figures~\ref{RKHS_fig} and~\ref{NN_fig} show that SRM attains in fact an optimal approximation-estimation error trade-off, thus validating our learning guarantee (see Remark~\ref{remark_SRM_guarantee}). This can be explained by the dependence of the penalties \eqref{RKHS_penalty}, \eqref{NN_penalty} on the \textit{learned} $k$-class predictors $\barf_k(\cdot)$ via the discretized term $M_{q(f_k)}$ (see Remark~\ref{rem_discret}).  

\section{Future Work}
Going forward, our paper opens up several research directions. First, in order to improve the statistical efficiency of our SRM scheme, we could exploit bounds for dependent data to analyze each trajectory. Moreover, an extension of our SRM method to nonlinear dynamical systems with additive noise and external inputs could be considered. Finally, it would be interesting to provide explicit SRM schemes for other model classes, such as Recurrent Neural Networks (RNNs).

\appendix

\subsection{Empirical Rademacher Complexity}\label{appendix_A}
The Rademacher complexity captures the richness of a function class by measuring its ability to fit symmetric random noise. Below we present the formal definition of the empirical Rademacher complexity.

\begin{definition}\label{def1}
\cite[Definition 3.1]{Mohri}
Let $\calG$ be a family of functions from $\calZ$ to $[a,b]$ and $S:=\{z_i\}_{i=1}^N$ a fixed sample of size $N$ with elements in $\calZ$. Then, the empirical Rademacher complexity of $\calG$ with respect to the sample $S$ is defined as:
\begin{equation*}\label{empirical_Rademacher}
    \hat{\calR}_S(\calG) = \Exp_{\boldsymbol{\sigma}}\left[\sup_{g\in\calG}\frac{1}{N}\sum_{i=1}^N\sigma_i g(z_i)\right],
\end{equation*}
where $\boldsymbol{\sigma}=\begin{bmatrix}\sigma_1,\ldots,\sigma_N\end{bmatrix}^\intercal$, with $\sigma_i$'s independent uniform random variables taking values in $\{-1,+1\}$. The random variables $\sigma_i$ are called Rademacher variables.
 \end{definition}
\noindent Notice that the empirical Rademacher complexity is a data-dependent class complexity measure, as it relies on the given sample $S$.

\subsection{Proofs}\label{appendix_B}
\subsubsection{Proof of Theorem~\ref{th1}}
From \cite[Theorem 3.5]{Mohri} we deduce that for each $k$ and every $\delta'\in(0,1)$, with probability at least $1-\delta'$ over the draw of $S\sim\calD^N$:
\begin{equation}\label{th1_3}
    \forall h\in\calH_k: \er_{\calD}^{0-1}[h]-\hat{\er}_S^{0-1}[h]\leq\epsilon_k(S,2\delta')
\end{equation}
holds, with $\epsilon_k(\cdot,\cdot)$ as given in the theorem statement. Similarly to \eqref{th1_3}, we can easily prove that for each $k$ and every $\delta'\in(0,1)$, with probability at least $1-\delta'$ over the draw of $S\sim\calD^N$:
\begin{equation}\label{th1_4}
    \forall h\in\calH_k: \hat{\er}_S^{0-1}[h]-\er_{\calD}^{0-1}[h]\leq\epsilon_k(S,2\delta')
\end{equation}
holds, with $\epsilon_k(\cdot,\cdot)$ as given in the theorem statement. Let $\calC_k(\delta')$ and $\calE_k(\delta')$ denote the sets of samples $S\sim\calD^N$ that satisfy \eqref{th1_3} and \eqref{th1_4}, respectively. Then, by De Morgan's law and union bound, for each $k$, we have:
\begin{align*}
&\Prob_{S}\left[S\in\calC_k(\delta')\cap \calE_k(\delta')\right]\\
&=1-\Prob_{S}\left[S\in\calC^c_k(\delta')\cup \calE^c_k(\delta')\right]\\
&\geq1-\Prob_{S}\left[S\in\calC^c_k(\delta')\right]-\Prob_{S}\left[S\in\calE^c_k(\delta')\right]\\
&\geq1-2\delta'.
\end{align*}
Hence, setting $\delta'=\delta/2$, we deduce that with probability at least $1-\delta$ over the draw of $S\sim\calD^N$, \eqref{ERM_UC_bound} holds with $\epsilon_k(\cdot,\cdot)$ as given in the theorem statement.

Fix any $h'\in\calH$. Then, by De Morgan's law and union bound, for every $\delta'\in(0,1)$ we can write:
\begin{align*}
&\Prob_{S}\Bigg[\Bigg\{\sup_{h\in\calH}\left(\er_{\calD}^{0-1}[h]-\hat{\er}_S^{0-1}[h]-\epsilon_{k(h)}(S,2\delta')\right)\leq0\Bigg\}\\
&\hspace{1cm}\cap\left\{S\in\calE_{k(h')}(\delta')\right\}\Bigg]\\
&=\Prob_{S}\left[S\in\left(\bigcap_{k=1}^{K}\calC_k(\delta')\right)\cap\calE_{k(h')}(\delta')\right]\\
&=1-\Prob_{S}\left[S\in\left(\bigcup_{k=1}^{K}\calC_k^c(\delta')\right)\cup\calE_{k(h')}^c(\delta')\right]\\
&\geq1-\sum_{k=1}^{K}\Prob_{S}\left[S\in\calC_k^c(\delta')\right]-\Prob_{S}\left[S\in\calE_{k(h')}^c(\delta')\right]\\
&\geq1-(K+1)\delta'.
\end{align*}
Therefore, setting $\delta'=\delta/(K+1)$, we conclude that with probability at least $1-\delta$ over the draw of $S\sim\calD^N$, the following bounds (simultaneously) hold:
\begin{align}\label{th1_5}  \forall h\in\calH&: \er_{\calD}^{0-1}[h]-\hat{\er}_S^{0-1}[h]\leq\epsilon_{k(h)}\left(S,\frac{2\delta}{K+1}\right)\\
    \label{th1_6}    \forall h\in\calH_{k(h')}&: \hat{\er}_S^{0-1}[h]-\er_{\calD}^{0-1}[h]\leq\epsilon_{k(h')}\left(S,\frac{2\delta}{K+1}\right).
\end{align}
Following similar arguments to \cite[Theorem 4.2]{Mohri}, we have:
\begin{align*}
&\er_{\calD}^{0-1}[h_S^{\SRM}]-\er_{\calD}^{0-1}[h']-2\hat{\calR}_S(\calH_{k(h')})\\
&\tleq{\eqref{SRM_classifier}}\left(\er_{\calD}^{0-1}[h_S^{\SRM}]-\hat{\er}_S^{0-1}[h_S^{\SRM}]-\hat{\calR}_S(\calH_{k(h_S^{\SRM})})\right)\\
&+\left(\hat{\er}_S^{0-1}[h']-\er_{\calD}^{0-1}[h']-\hat{\calR}_S(\calH_{k(h')})\right)\\
&\leq\sup_{h\in\calH}\left(\er_{\calD}^{0-1}[h]-\hat{\er}_S^{0-1}[h]-\hat{\calR}_S(\calH_{k(h)})\right)\\
&+\sup_{h\in\calH_{k(h')}}\left(\hat{\er}_S^{0-1}[h]-\er_{\calD}^{0-1}[h]-\hat{\calR}_S(\calH_{k(h')})\right)\\
&\leq6\sqrt{\frac{\log\left((2(K+1))/\delta\right)}{2N}},
\end{align*}
with probability at least $1-\delta$ over the draw of $S\sim\calD^N$, given \eqref{th1_5} and \eqref{th1_6}. Consequently, setting $h'$ as the function that attains the minimum on the right-hand side of \eqref{SRM_guarantee}, we deduce that with probability at least $1-\delta$ over the draw of $S\sim\calD^N$, \eqref{SRM_guarantee} holds with $\epsilon_k(\cdot,\cdot)$ as given in the theorem statement.\\
\qed

\subsubsection{Proof of Theorem~\ref{th2}}
Let us first note that since the set $Z$ is compact, by Assumption~\ref{ass_loss} we deduce that $\ell(\cdot)$ is Lipschitz continuous on $Z$ and thus a corresponding Lipschitz constant $L>0$ is guaranteed to exist. Moreover, for each $k$, let us define the loss class $\calL_k = \Big\{x_{0:T}(\xi)\mapsto\frac{1}{T}\sum_{t=0}^{T-1}\ell(\thickbar{f}(x_t(\xi))-x_{t+1}(\xi))\,\Big|\, f\in\calF_k\Big\}$, where $\xi\sim\calD$. We will prove that for each $k$:
\begin{equation}\label{th2_1}
    \hat{\calR}_S(\calL_k)\leq\frac{\sqrt{2}L}{T}\sum_{t=0}^{T-1}\sum_{j=1}^n\hat{\calR}_{S_t}(\thickbar{\calF}_{k,j}).
\end{equation}
Using the triangle inequality, Assumption~\ref{ass_sys} and \eqref{normalized_predictor}, we can easily show that for each $k$:
\begin{equation}\label{th2_0}
    \sup_{f\in\calF_k,\xi\sim\calD}\norm{\barf(x_t(\xi))-x_{t+1}(\xi)}_2\leq2B.
\end{equation}
Let $\bvarepsilon=\begin{bmatrix}\bvarepsilon_1,\ldots,\bvarepsilon_n\end{bmatrix}$, where for each $j$, $\bvarepsilon_j=\begin{bmatrix}\varepsilon_{1j},\ldots,\varepsilon_{Nj}\end{bmatrix}^\intercal$, with $\varepsilon_{ij}$'s independent Rademacher variables. Employing \cite[Lemma 5]{Cortes2016} as well as subadditivity of supremum and linearity of expectation, we can write:
\begin{align*}
&\hat{\calR}_S(\calL_k)=\Exp_{\bsigma}\Bigg[\sup_{f\in\calF_k}\frac{1}{N}\sum_{i=1}^N\sigma_i\Bigg(\frac{1}{T}\sum_{t=0}^{T-1}\ell(\thickbar{f}(x_t(\xi_i))\\
&\hspace{2.4cm}-x_{t+1}(\xi_i))\Bigg)\Bigg]\\
&\leq\frac{1}{T}\sum_{t=0}^{T-1}\Exp_{\bsigma}\left[\sup_{f\in\calF_k}\frac{1}{N}\sum_{i=1}^N\sigma_i\ell(\thickbar{f}(x_t(\xi_i))-x_{t+1}(\xi_i))\right]\\
&\leq\frac{\sqrt{2}L}{T}\sum_{t=0}^{T-1}\Exp_{\bvarepsilon}\left[\sup_{f\in\calF_k}\frac{1}{N}\sum_{i=1}^N\sum_{j=1}^n\varepsilon_{ij}\barf_j(x_t(\xi_i))\right]\\
&\leq \frac{\sqrt{2}L}{T}\sum_{t=0}^{T-1}\sum_{j=1}^n\Exp_{\bvarepsilon_j}\left[\sup_{\barf_j\in\thickbar{\calF}_{k,j}}\frac{1}{N}\sum_{i=1}^N\varepsilon_{ij}\barf_j(x_t(\xi_i))\right]\\
&=\frac{\sqrt{2}L}{T}\sum_{t=0}^{T-1}\sum_{j=1}^n\hat{\calR}_{S_t}(\thickbar{\calF}_{k,j}),
\end{align*}
which completes the proof of \eqref{th2_1}. Given the assumption that $\ell(0)=0$ (see Assumption~\ref{ass_loss}), by definition of $L$ and \eqref{th2_0} we conclude that for each $k$:
\begin{align}\label{th2_2}
\sup_{f\in\calF_k,\xi\sim\calD}\frac{1}{T}\sum_{t=0}^{T-1}\ell(\thickbar{f}(x_t(\xi))-x_{t+1}(\xi))\leq2LB.
\end{align}
From \eqref{th2_1}, \eqref{th2_2} and a straightforward extension of \cite[Theorem 3.3]{Mohri} to classes of functions bounded in $[0,2LB]$ instead of $[0,1]$, we deduce that for each $k$ and every $\delta'\in(0,1)$, with probability at least $1-\delta'$ over the draw of $S_0\sim\calD^N$:
\begin{equation}\label{th2_3}
    \forall f\in\calF_k: \er_\calD[\barf]-\hat{\er}_S[\barf]\leq\epsilon_k(S,2\delta')
\end{equation}
holds, with $\epsilon_k(\cdot,\cdot)$ as given in the theorem statement. Similarly to \eqref{th2_3}, we can easily prove that for each $k$ and every $\delta'\in(0,1)$, with probability at least $1-\delta'$ over the draw of $S_0\sim\calD^N$:
\begin{equation}\label{th2_4}
    \forall f\in\calF_k: \hat{\er}_S[\barf]-\er_\calD[\barf]\leq\epsilon_k(S,2\delta')
\end{equation}
holds, with $\epsilon_k(\cdot,\cdot)$ as given in the theorem statement. Let $\calC_k(\delta')$ and $\calE_k(\delta')$ denote the sets of samples $S_0\sim\calD^N$ that satisfy \eqref{th2_3} and \eqref{th2_4}, respectively. Then, by De Morgan's law and union bound, for each $k$, we have:
\begin{align*}
&\Prob_{S_0}\left[S_0\in\calC_k(\delta')\cap \calE_k(\delta')\right]\\
&=1-\Prob_{S_0}\left[S_0\in\calC^c_k(\delta')\cup \calE^c_k(\delta')\right]\\
&\geq1-\Prob_{S_0}\left[S_0\in\calC^c_k(\delta')\right]-\Prob_{S_0}\left[S_0\in\calE^c_k(\delta')\right]\\
&\geq1-2\delta'.
\end{align*}
Hence, setting $\delta'=\delta/2$, we deduce that with probability at least $1-\delta$ over the draw of $S_0\sim\calD^N$, \eqref{ERM_UC_bound1} holds with $\epsilon_k(\cdot,\cdot)$ as given in the theorem statement.

Fix any $f'\in\calF$. Then, by De Morgan's law and union bound, for every $\delta'\in(0,1)$ we can write:
\begin{align*}
&\Prob_{S_0}\Bigg[\Bigg\{\sup_{f\in\calF}\left(\er_\calD[\barf]-\hat{\er}_S[\barf]-\epsilon_{k(f)}(S,2\delta')\right)\leq0\Bigg\}\\
&\hspace{1cm}\cap\left\{S_0\in\calE_{k(f')}(\delta')\right\}\Bigg]\\
&=\Prob_{S_0}\left[S_0\in\left(\bigcap_{k=1}^{K}\calC_k(\delta')\right)\cap\calE_{k(f')}(\delta')\right]\\
&=1-\Prob_{S_0}\left[S_0\in\left(\bigcup_{k=1}^{K}\calC_k^c(\delta')\right)\cup\calE_{k(f')}^c(\delta')\right]\\
&\geq1-\sum_{k=1}^{K}\Prob_{S_0}\left[S_0\in\calC_k^c(\delta')\right]-\Prob_{S_0}\left[S_0\in\calE_{k(f')}^c(\delta')\right]\\
&\geq1-(K+1)\delta'.
\end{align*}
Therefore, setting $\delta'=\delta/(K+1)$, we conclude that with probability at least $1-\delta$ over the draw of $S_0\sim\calD^N$, the following bounds (simultaneously) hold:
\begin{align}\label{th2_5}  \forall f\in\calF&: \er_\calD[\barf]-\hat{\er}_S[\barf]\leq\epsilon_{k(f)}\left(S,\frac{2\delta}{K+1}\right)\\
    \label{th2_6}    \forall f\in\calF_{k(f')}&: \hat{\er}_S[\barf]-\er_\calD[\barf]\leq\epsilon_{k(f')}\left(S,\frac{2\delta}{K+1}\right).
\end{align}
Following similar arguments to \cite[Theorem 4.2]{Mohri}, we have:
\begin{align*}
&\er_\calD[\barf_S^{\SRM}]-\er_\calD[\barf']-2r_{k(f')}\left(S\right)\\
&\tleq{\eqref{SRM_predictor}}\left(\er_\calD[\barf_S^{\SRM}]-\hat{\er}_S[\barf_S^{\SRM}]-r_{k(f_S^{\SRM})}\left(S\right)\right)\\
&+\left(\hat{\er}_S[\barf']-\er_\calD[\barf']-r_{k(f')}\left(S\right)\right)\\
&\leq\sup_{f\in\calF}\left(\er_\calD[\barf]-\hat{\er}_S[\barf]-r_{k(f)}\left(S\right)\right)\\
&+\sup_{f\in\calF_{k(f')}}\left(\hat{\er}_S[\barf]-\er_\calD[\barf]-r_{k(f')}\left(S\right)\right)\\
&\leq12LB\sqrt{\frac{\log\left((2(K+1))/\delta\right)}{2N}},
\end{align*}
with probability at least $1-\delta$ over the draw of $S_0\sim\calD^N$, given \eqref{th2_5} and \eqref{th2_6}. Consequently, setting $f'$ as the function that attains the minimum on the right-hand side of \eqref{SRM_guarantee1}, we deduce that with probability at least $1-\delta$ over the draw of $S_0\sim\calD^N$, \eqref{SRM_guarantee1} holds with $\epsilon_k(\cdot,\cdot)$ as given in the theorem statement.\\
\qed

\subsubsection{Proof of Theorem~\ref{th3}}
Given the fact that $M_Q=\max_{k,f\in\calF_k}\max_j\norm{f_j}_{\setH_k}$, for each $k$, $\calF_k$ can be written as $\calF_k=\bigcup_{q=1}^Q\calF_{kq}$, where $\calF_{kq}=\Big\{f\in\calF_k\,\Big|\max_j\norm{f_j}_{\setH_k}\leq M_q\Big\}$, $\forall q$. For each $k$ and $q$, let us define $\epsilon_{kq}(S,\delta'')=R_{kq}(S)+6LB\sqrt{\log(4/\delta'')/(2N)}$, $\forall\delta''\in(0,1)$, where:
\begin{equation*}\label{th3_1}
    R_{kq}(S)=\frac{2\sqrt{2}LnM_q}{TN}\sum\limits_{t=0}^{T-1}\sqrt{\sum\limits_{i=1}^N\kappa_{\theta_k}\left(x_t(\xi_i),x_t(\xi_i)\right)}.
\end{equation*}
From \cite[Theorem 6.12]{Mohri} we deduce that for each $k$ and $q$, the following penalty:
\begin{equation*}
    r_{kq}(S)=\frac{2\sqrt{2}L}{T}\sum_{t=0}^{T-1}\sum_{j=1}^n\hat{\calR}_{S_t}(\thickbar{\calF}_{kq,j}),
\end{equation*}
where $\thickbar{\calF}_{kq,j}=\left\{j\text{-th component of }\barf(\cdot)\where f\in\calF_{kq}\right\}$ and $S_t=\{x_t(\xi_i)\}_{i=1}^N$,
can be upper-bounded by $R_{kq}(S)$. Employing the bounds $R_{kq}(S)$ instead of the penalties $r_{kq}(S)$ and following the proof of \eqref{th2_3} and \eqref{th2_4}, we can show that for each $k$ and $q$, and every $\delta''\in(0,1)$, each of the following:
\begin{align}\label{th3_2}
    \forall f\in\calF_{kq}&: \er_\calD[\barf]-\hat{\er}_S[\barf]\leq\epsilon_{kq}(S,2\delta'')\\\label{th3_3}
    \forall f\in\calF_{kq}&: \hat{\er}_S[\barf]-\er_\calD[\barf]\leq\epsilon_{kq}(S,2\delta'')
\end{align}
holds, with probability at least $1-\delta''$ over the draw of $S_0\sim\calD^N$.  Let $\calC_{kq}(\delta'')$ denote the set of samples $S_0\sim\calD^N$ that satisfy \eqref{th3_2} and set $\calC_k(\delta'')=\bigcap_{q=1}^Q\calC_{kq}(\delta'')$. Then, by union bound and De Morgan's law, for each $k$, we have:
\begin{align*}
&\Prob_{S_0}\left[S_0\in\calC_{k}(\delta'')\right]=1-\Prob_{S_0}\left[S_0\in\bigcup_{q=1}^Q\calC_{kq}^c(\delta'')\right]\\
&\geq1-\sum_{q=1}^Q\Prob_{S_0}\left[S_0\in\calC^c_{kq}(\delta'')\right]\geq1-Q\delta''.
\end{align*}
Therefore, for any $\delta'\in(0,1)$, setting $\delta''=\delta'/Q$, we deduce that with probability at least $1-\delta'$ over the draw of $S_0\sim\calD^N$:
\begin{align}\label{th3_4}
    \forall f\in\calF_k&: \er_\calD[\barf]-\hat{\er}_S[\barf]\leq\epsilon_k(S,2\delta')
\end{align}
holds, with $\epsilon_k(\cdot,\cdot)$ as given in the theorem statement. Similarly, from \eqref{th3_3} we can prove that for every $\delta'\in(0,1)$, with probability at least $1-\delta'$ over the draw of $S_0\sim\calD^N$:
\begin{align}\label{th3_5}
    \forall f\in\calF_k&: \hat{\er}_S[\barf]-\er_\calD[\barf]\leq\epsilon_k(S,2\delta')
\end{align}
holds, with $\epsilon_k(\cdot,\cdot)$ as given in the theorem statement. Hence, setting $\delta'=\delta/2$, from \eqref{th3_4} and \eqref{th3_5} we can easily deduce that with probability at least $1-\delta$ over the draw of $S_0\sim\calD^N$, \eqref{ERM_UC_bound1} holds with $\epsilon_k(\cdot,\cdot)$ as given in the theorem statement.

Fix any $f'\in\calF$. Notice that \eqref{th3_4} and \eqref{th3_5} have the same form as \eqref{th2_3} and \eqref{th2_4}, respectively, even though for a different definition of $\epsilon_k(\cdot,\cdot)$. Therefore, similarly to the proof of \eqref{th2_5} and \eqref{th2_6}, we can show that with probability at least $1-\delta$ over the draw of $S_0\sim\calD^N$, the following bounds (simultaneously) hold:
\begin{align}\label{th3_6}  \forall f\in\calF&: \er_\calD[\barf]-\hat{\er}_S[\barf]\leq\epsilon_{k(f)}\left(S,\frac{2\delta}{K+1}\right)\\
    \label{th3_7}    \forall f\in\calF_{k(f')}&: \hat{\er}_S[\barf]-\er_\calD[\barf]\leq\epsilon_{k(f')}\left(S,\frac{2\delta}{K+1}\right).
\end{align}
Following similar arguments to \cite[Theorem 4.2]{Mohri} and employing the definition of $\tilde{r}_{k}(S)$, we have:
\begin{align*}
&\er_\calD[\barf_S^{\SRM}]-\er_\calD[\barf']-2\tilde{r}_{k(f')}\left(S\right)\\
&\tleq{\eqref{SRM_predictor}}\er_\calD[\barf_S^{\SRM}]-\er_\calD[\barf']-2\tilde{r}_{k(f')}(S)-\big(\hat{\er}_S[\barf_S^{\SRM}]\\
&+r_{k(f_S^{\SRM})}\left(S\right)\big)+\left(\hat{\er}_S[\barf']+r_{k(f')}\left(S\right)\right)\\
&\leq\left(\er_\calD[\barf_S^{\SRM}]-\hat{\er}_S[\barf_S^{\SRM}]-\tilde{r}_{k(f_S^{\SRM})}\left(S\right)\right)\\
&+\left(\hat{\er}_S[\barf']-\er_\calD[\barf']-\tilde{r}_{k(f')}\left(S\right)\right)\\
&\leq\sup_{f\in\calF}\left(\er_\calD[\barf]-\hat{\er}_S[\barf]-\tilde{r}_{k(f)}\left(S\right)\right)\\
&+\sup_{f\in\calF_{k(f')}}\left(\hat{\er}_S[\barf]-\er_\calD[\barf]-\tilde{r}_{k(f')}\left(S\right)\right)\\
&\leq12LB\sqrt{\frac{\log\left((2(K+1)Q)/\delta\right)}{2N}},
\end{align*}
with probability at least $1-\delta$ over the draw of $S_0\sim\calD^N$, given \eqref{th3_6} and \eqref{th3_7}. Consequently, setting $f'$ as the function that attains the minimum on the right-hand side of \eqref{SRM_guarantee1}, we deduce that with probability at least $1-\delta$ over the draw of $S_0\sim\calD^N$, \eqref{SRM_guarantee1} holds with $\epsilon_k(\cdot,\cdot)$ as given in the theorem statement.\\
\qed

\subsubsection{Proof of Theorem~\ref{th4}}
Given the fact that $M_Q=\max_{k,f\in\calF_k}\max_d\norm{W_f^{(d)}}_{\F}$, for each $k$, $\calF_k$ can be written as $\calF_k=\bigcup_{q=1}^Q\calF_{kq}$, where $\calF_{kq}=\Big\{f\in\calF_k\,\Big|\max_d\norm{W_f^{(d)}}_{\F}\leq M_q\Big\}$, $\forall q$. For each $k$ and $q$, let us define $\epsilon_{kq}(S,\delta'')=R_{kq}(S)+6LB\sqrt{\log(4/\delta'')/(2N)}$, $\forall\delta''\in(0,1)$, where:
\begin{equation*}\label{th4_1}
R_{kq}(S)=\frac{LnM_q^{D_k}a(D_k)}{TN}\sum_{t=0}^{T-1}\sqrt{\sum_{i=1}^N\left(\norm{x_t(\xi_i))}_2^2+1\right)},
\end{equation*}
with $a(\cdot)$ as given in the theorem statement.
Employing \cite[Theorem 1]{Golowich2019} and performing straightforward algebraic manipulations, we deduce that for each $k$ and $q$, the following penalty:
\begin{equation*}
    r_{kq}(S)=\frac{2\sqrt{2}L}{T}\sum_{t=0}^{T-1}\sum_{j=1}^n\hat{\calR}_{S_t}(\thickbar{\calF}_{kq,j}),
\end{equation*}
where $\thickbar{\calF}_{kq,j}=\left\{j\text{-th component of }\barf(\cdot)\where f\in\calF_{kq}\right\}$ and $S_t=\{x_t(\xi_i)\}_{i=1}^N$,
can be upper-bounded by $R_{kq}(S)$. Using the bounds $R_{kq}(S)$ instead of the penalties $r_{kq}(S)$ and following the proof of \eqref{th2_3} and \eqref{th2_4}, we can show that for each $k$ and $q$, and every $\delta''\in(0,1)$, each of the following:
\begin{align}\label{th4_2}
    \forall f\in\calF_{kq}&: \er_\calD[\barf]-\hat{\er}_S[\barf]\leq\epsilon_{kq}(S,2\delta'')\\\label{th4_3}
    \forall f\in\calF_{kq}&: \hat{\er}_S[\barf]-\er_\calD[\barf]\leq\epsilon_{kq}(S,2\delta'')
\end{align}
holds, with probability at least $1-\delta''$ over the draw of $S_0\sim\calD^N$.  Let $\calC_{kq}(\delta'')$ denote the set of samples $S_0\sim\calD^N$ that satisfy \eqref{th4_2} and set $\calC_k(\delta'')=\bigcap_{q=1}^Q\calC_{kq}(\delta'')$. Then, by union bound and De Morgan's law, for each $k$, we have:
\begin{align*}
&\Prob_{S_0}\left[S_0\in\calC_{k}(\delta'')\right]=1-\Prob_{S_0}\left[S_0\in\bigcup_{q=1}^Q\calC_{kq}^c(\delta'')\right]\\
&\geq1-\sum_{q=1}^Q\Prob_{S_0}\left[S_0\in\calC^c_{kq}(\delta'')\right]\geq1-Q\delta''.
\end{align*}
Therefore, for any $\delta'\in(0,1)$, setting $\delta''=\delta'/Q$, we deduce that with probability at least $1-\delta'$ over the draw of $S_0\sim\calD^N$:
\begin{align}\label{th4_4}
    \forall f\in\calF_k&: \er_\calD[\barf]-\hat{\er}_S[\barf]\leq\epsilon_k(S,2\delta')
\end{align}
holds, with $\epsilon_k(\cdot,\cdot)$ as given in the theorem statement. Similarly, from \eqref{th4_3} we can prove that for every $\delta'\in(0,1)$, with probability at least $1-\delta'$ over the draw of $S_0\sim\calD^N$:
\begin{align}\label{th4_5}
    \forall f\in\calF_k&: \hat{\er}_S[\barf]-\er_\calD[\barf]\leq\epsilon_k(S,2\delta')
\end{align}
holds, with $\epsilon_k(\cdot,\cdot)$ as given in the theorem statement. Hence, setting $\delta'=\delta/2$, from \eqref{th4_4} and \eqref{th4_5} we can easily deduce that with probability at least $1-\delta$ over the draw of $S_0\sim\calD^N$, \eqref{ERM_UC_bound1} holds with $\epsilon_k(\cdot,\cdot)$ as given in the theorem statement.

Fix any $f'\in\calF$. Notice that \eqref{th4_4} and \eqref{th4_5} have the same form as \eqref{th2_3} and \eqref{th2_4}, respectively, even though for a different definition of $\epsilon_k(\cdot,\cdot)$. Therefore, similarly to the proof of \eqref{th2_5} and \eqref{th2_6}, we can show that with probability at least $1-\delta$ over the draw of $S_0\sim\calD^N$, the following bounds (simultaneously) hold:
\begin{align}\label{th4_6}  \forall f\in\calF&: \er_\calD[\barf]-\hat{\er}_S[\barf]\leq\epsilon_{k(f)}\left(S,\frac{2\delta}{K+1}\right)\\
    \label{th4_7}    \forall f\in\calF_{k(f')}&: \hat{\er}_S[\barf]-\er_\calD[\barf]\leq\epsilon_{k(f')}\left(S,\frac{2\delta}{K+1}\right).
\end{align}
Following similar arguments to \cite[Theorem 4.2]{Mohri} and employing the definition of $\tilde{r}_{k}(S)$, we have:
\begin{align*}
&\er_\calD[\barf_S^{\SRM}]-\er_\calD[\barf']-2\tilde{r}_{k(f')}\left(S\right)\\
&\tleq{\eqref{SRM_predictor}}\er_\calD[\barf_S^{\SRM}]-\er_\calD[\barf']-2\tilde{r}_{k(f')}(S)-\big(\hat{\er}_S[\barf_S^{\SRM}]\\
&+r_{k(f_S^{\SRM})}\left(S\right)\big)+\left(\hat{\er}_S[\barf']+r_{k(f')}\left(S\right)\right)\\
&\leq\left(\er_\calD[\barf_S^{\SRM}]-\hat{\er}_S[\barf_S^{\SRM}]-\tilde{r}_{k(f_S^{\SRM})}\left(S\right)\right)\\
&+\left(\hat{\er}_S[\barf']-\er_\calD[\barf']-\tilde{r}_{k(f')}\left(S\right)\right)\\
&\leq\sup_{f\in\calF}\left(\er_\calD[\barf]-\hat{\er}_S[\barf]-\tilde{r}_{k(f)}\left(S\right)\right)\\
&+\sup_{f\in\calF_{k(f')}}\left(\hat{\er}_S[\barf]-\er_\calD[\barf]-\tilde{r}_{k(f')}\left(S\right)\right)\\
&\leq12LB\sqrt{\frac{\log\left((2(K+1)Q)/\delta\right)}{2N}},
\end{align*}
with probability at least $1-\delta$ over the draw of $S_0\sim\calD^N$, given \eqref{th4_6} and \eqref{th4_7}. Consequently, setting $f'$ as the function that attains the minimum on the right-hand side of \eqref{SRM_guarantee1}, we deduce that with probability at least $1-\delta$ over the draw of $S_0\sim\calD^N$, \eqref{SRM_guarantee1} holds with $\epsilon_k(\cdot,\cdot)$ as given in the theorem statement.\\
\qed

\subsection{Implementation of RKHS-based SRM for Nonlinear Dynamics}\label{appendix_C}
In this subsection, we provide a practical method for implementing the RKHS-based SRM scheme proposed in Theorem~\ref{th3}. The challenge lies in the fact that the corresponding optimization problem in \eqref{SRM_predictor} is defined over a possibly infinite-dimensional space of functions. To address this issue, we transform the original problem into a reduced problem with a finite number of real optimization variables. Herein, by reduced problem we mean a problem that admits at least one solution which is also a solution of the original problem.

Following the proof of \cite[Theorem 6.11]{Mohri}, for each $k$, we define $\tilde{\setH}_{k}=\textup{span}(\{k_{\theta_k}(x_t(\xi_i),\cdot):i=1,\ldots,N,\,t=0,\ldots,T-1\})$. Note that any $f:=\begin{bmatrix}f_1,\ldots,f_n\end{bmatrix}^{\intercal}\in\calF_k$ admits a decomposition of the form $f=f^1+f^{\perp}$, where $f^1:=\begin{bmatrix}f^1_{1},\ldots,f^1_{n}\end{bmatrix}^{\intercal}$ and $f^{\perp}:=\begin{bmatrix}f_1^{\perp},\ldots,f_n^{\perp}\end{bmatrix}^{\intercal}$
result from the component-wise decompositions $f_j=f^1_{j}+f_j^{\perp}$
according to $\setH_k=\tilde{\setH}_{k}\oplus\tilde{\setH}_{k}^{\perp}$. By the reproducing property and the definition of $f^1$, for each $j$, we can write:
\begin{align*}
    f_j(x_t(\xi_i))&=\langle f_j,k_{\theta_k}(x_t(\xi_i),\cdot)\rangle\\
    &=\langle f^1_j,k_{\theta_k}(x_t(\xi_i),\cdot)\rangle=f^1_j(x_t(\xi_i)),
\end{align*}
for all $i=1,\ldots,N$ and $t=0,\ldots,T-1$. Hence, we deduce that $\hat{\er}_S[\barf]=\hat{\er}_S[\thickbar{f^1}]$, $\forall f\in\calF_k$. Consequently, for each $k$, the optimization problem:
\begin{equation}\label{RKHS_original}
\min_{f\in\calF_k}\hat{\er}_S[\barf]
\end{equation}
admits a solution of the form:
\begin{equation*}
    f(x)=\sum_{i=1}^N\sum_{t=0}^{T-1}\balpha^{(i,t)}\kappa_{\theta_k}(x_t(\xi_i),x),
\end{equation*}
where $\balpha^{(i,t)}=\begin{bmatrix}\alpha_1^{(i,t)},\ldots,\alpha_n^{(i,t)}\end{bmatrix}^{\intercal}\in\setR^n$, $\forall i$, $t$. Note that for every function  $f:=\begin{bmatrix}f_1,\ldots,f_n\end{bmatrix}^\intercal$ of the above form, we have:
\begin{equation}\label{RKHS_norm}
\norm{f_j}_{\setH_k}=\sqrt{\sum_{i,i',t,t'}\alpha_j^{(i,t)}\alpha_j^{(i',t')}\kappa_{\theta_k}(x_t(\xi_i),x_{t'}(\xi_{i'}))},
\end{equation}
for all $j=1,\ldots,n$. Therefore, for each $k$, we can reduce problem \eqref{RKHS_original} into the problem:
\begin{equation}\label{reduced_problem}
\min_{f\in\widetilde{\calF}_k}\hat{\er}_S[\barf],
\end{equation}
where:  
\begin{align*}
    &\widetilde{\calF}_k=\Bigg\{x\mapsto\sum_{i=1}^N\sum_{t=0}^{T-1}\balpha^{(i,t)}\kappa_{\theta_k}(x_t(\xi_i),x)\,\Bigg|\nonumber\\&\sum_{i,i',t,t'}\alpha_j^{(i,t)}\alpha_j^{(i',t')}\kappa_{\theta_k}(x_t(\xi_i),x_{t'}(\xi_{i'}))\leq \calB_k^2,\forall j\Bigg\}.
\end{align*}
Note that the optimization problem \eqref{reduced_problem} is tractable as it is defined over the finite-dimensional space of $NT$ vectors $\balpha^{(i,t)}\in\setR^n$.
Hence, we conclude that the RKHS-based SRM predictor can be computed as in \eqref{SRM_predictor} by simply replacing the classes $\calF_k$ with the classes $\widetilde{\calF}_k$. In this case, for each $k$, the learned model is given by $f_k=\argmin_{f\in\tilde{\calF}_K}\hat{\er}_S[\barf]$ and the value of $M_{q(f_k)}$ in the penalty \eqref{RKHS_penalty} can be computed using \eqref{RKHS_norm}.

\bibliographystyle{IEEEtran} 
\bibliography{arxiv_version}

\begin{thebibliography}{10}
\providecommand{\url}[1]{#1}
\csname url@rmstyle\endcsname
\providecommand{\newblock}{\relax}
\providecommand{\bibinfo}[2]{#2}
\providecommand\BIBentrySTDinterwordspacing{\spaceskip=0pt\relax}
\providecommand\BIBentryALTinterwordstretchfactor{4}
\providecommand\BIBentryALTinterwordspacing{\spaceskip=\fontdimen2\font plus
\BIBentryALTinterwordstretchfactor\fontdimen3\font minus \fontdimen4\font\relax}
\providecommand\BIBforeignlanguage[2]{{%
\expandafter\ifx\csname l@#1\endcsname\relax
\typeout{** WARNING: IEEEtran.bst: No hyphenation pattern has been}%
\typeout{** loaded for the language `#1'. Using the pattern for}%
\typeout{** the default language instead.}%
\else
\language=\csname l@#1\endcsname
\fi
#2}}

\bibitem{Anderson1995}
J.~D. Anderson and J.~Wendt, \emph{Computational fluid dynamics}.\hskip 1em plus 0.5em minus 0.4em\relax Springer, 1995, vol. 206.

\bibitem{Rajkomar2014}
A.~Rajkomar, J.~Dean, and I.~Kohane, ``Machine learning in medicine,'' \emph{New England Journal of Medicine}, vol. 380, no.~14, pp. 1347--1358, 2019.

\bibitem{Day1994}
R.~H. Day, \emph{Complex Economic Dynamics - Vol. 1: An Introduction to Dynamical Systems and Market Mechanisms}, ser. MIT Press Books.\hskip 1em plus 0.5em minus 0.4em\relax The MIT Press, 1994, vol.~1.

\bibitem{Ghahramani1998}
Z.~Ghahramani and S.~Roweis, ``Learning nonlinear dynamical systems using an em algorithm,'' \emph{Advances in neural information processing systems}, vol.~11, 1998.

\bibitem{Langford2009}
J.~Langford, R.~Salakhutdinov, and T.~Zhang, ``Learning nonlinear dynamic models,'' in \emph{Proceedings of the 26th Annual International Conference on Machine Learning}, 2009, pp. 593--600.

\bibitem{Khansari2011}
S.~M. Khansari-Zadeh and A.~Billard, ``Learning stable nonlinear dynamical systems with gaussian mixture models,'' \emph{IEEE Transactions on Robotics}, vol.~27, no.~5, pp. 943--957, 2011.

\bibitem{Pillonetto2014}
G.~Pillonetto, F.~Dinuzzo, T.~Chen, G.~De~Nicolao, and L.~Ljung, ``Kernel methods in system identification, machine learning and function estimation: A survey,'' \emph{Automatica}, vol.~50, no.~3, pp. 657--682, 2014.

\bibitem{Svensson2017}
A.~Svensson and T.~B. Sch{\"o}n, ``A flexible state--space model for learning nonlinear dynamical systems,'' \emph{Automatica}, vol.~80, pp. 189--199, 2017.

\bibitem{Neumann2013}
K.~Neumann, A.~Lemme, and J.~J. Steil, ``Neural learning of stable dynamical systems based on data-driven lyapunov candidates,'' in \emph{2013 IEEE/RSJ International Conference on Intelligent Robots and Systems}, 2013, pp. 1216--1222.

\bibitem{Raissi2018}
M.~Raissi, P.~Perdikaris, and G.~E. Karniadakis, ``Multistep neural networks for data-driven discovery of nonlinear dynamical systems,'' \emph{arXiv preprint arXiv:1801.01236}, 2018.

\bibitem{Teng2019}
Q.~Teng and L.~Zhang, ``Data driven nonlinear dynamical systems identification using multi-step cldnn,'' \emph{AIP Advances}, vol.~9, no.~8, p. 085311, 2019.

\bibitem{Qin2021}
T.~Qin, Z.~Chen, J.~D. Jakeman, and D.~Xiu, ``Deep learning of parameterized equations with applications to uncertainty quantification,'' \emph{International Journal for Uncertainty Quantification}, vol.~11, no.~2, 2021.

\bibitem{Foster2020}
D.~Foster, T.~Sarkar, and A.~Rakhlin, ``Learning nonlinear dynamical systems from a single trajectory,'' in \emph{Learning for Dynamics and Control}, 2020, pp. 851--861.

\bibitem{Singh2021}
S.~Singh, S.~M. Richards, V.~Sindhwani, J.-J.~E. Slotine, and M.~Pavone, ``Learning stabilizable nonlinear dynamics with contraction-based regularization,'' \emph{The International Journal of Robotics Research}, vol.~40, no. 10-11, pp. 1123--1150, 2021.

\bibitem{Brunton2016}
S.~L. Brunton, J.~L. Proctor, and J.~N. Kutz, ``Discovering governing equations from data by sparse identification of nonlinear dynamical systems,'' \emph{Proceedings of the national academy of sciences}, vol. 113, no.~15, pp. 3932--3937, 2016.

\bibitem{Mangan2017}
N.~M. Mangan, J.~N. Kutz, S.~L. Brunton, and J.~L. Proctor, ``Model selection for dynamical systems via sparse regression and information criteria,'' \emph{Proceedings of the Royal Society A: Mathematical, Physical and Engineering Sciences}, vol. 473, no. 2204, p. 20170009, 2017.

\bibitem{Champion2019}
K.~Champion, B.~Lusch, J.~N. Kutz, and S.~L. Brunton, ``Data-driven discovery of coordinates and governing equations,'' \emph{Proceedings of the National Academy of Sciences}, vol. 116, no.~45, pp. 22\,445--22\,451, 2019.

\bibitem{Lusch2018}
B.~Lusch, J.~N. Kutz, and S.~L. Brunton, ``Deep learning for universal linear embeddings of nonlinear dynamics,'' \emph{Nature communications}, vol.~9, no.~1, p. 4950, 2018.

\bibitem{Kaiser2021}
E.~Kaiser, J.~N. Kutz, and S.~L. Brunton, ``Data-driven discovery of koopman eigenfunctions for control,'' \emph{Machine Learning: Science and Technology}, vol.~2, no.~3, p. 035023, 2021.

\bibitem{Lugosi1996}
G.~Lugosi and K.~Zeger, ``Concept learning using complexity regularization,'' \emph{IEEE Transactions on Information Theory}, vol.~42, no.~1, pp. 48--54, 1996.

\bibitem{Koltchinskii2001}
V.~Koltchinskii, ``Rademacher penalties and structural risk minimization,'' \emph{IEEE Transactions on Information Theory}, vol.~47, no.~5, pp. 1902--1914, 2001.

\bibitem{Shawe1998}
J.~Shawe-Taylor, P.~L. Bartlett, R.~C. Williamson, and M.~Anthony, ``Structural risk minimization over data-dependent hierarchies,'' \emph{IEEE transactions on Information Theory}, vol.~44, no.~5, pp. 1926--1940, 1998.

\bibitem{Gyurik2023}
C.~Gyurik, V.~Dunjko, \emph{et~al.}, ``Structural risk minimization for quantum linear classifiers,'' \emph{Quantum}, vol.~7, p. 893, 2023.

\bibitem{Liu2020structural}
J.~Liu, M.~Bai, N.~Jiang, and D.~Yu, ``Structural risk minimization of rough set-based classifier,'' \emph{Soft Computing}, vol.~24, pp. 2049--2066, 2020.

\bibitem{Massucci2020}
L.~Massucci, F.~Lauer, and M.~Gilson, ``Structural risk minimization for switched system identification,'' in \emph{2020 59th IEEE Conference on Decision and Control (CDC)}, 2020, pp. 1002--1007.

\bibitem{Meir1997}
R.~Meir, ``Structural risk minimization for nonparametric time series prediction,'' \emph{Advances in Neural Information Processing Systems}, vol.~10, 1997.

\bibitem{Mohri}
M.~Mohri, A.~Rostamizadeh, and A.~Talwalkar, \emph{Foundations of Machine Learning}.\hskip 1em plus 0.5em minus 0.4em\relax {MIT} Press, 2nd edition, 2018.

\bibitem{ShalevShwartz}
S.~Shalev-Shwartz and S.~Ben-David, \emph{Understanding Machine Learning: From Theory to Algorithms}.\hskip 1em plus 0.5em minus 0.4em\relax Cambridge University Press, 2014.

\bibitem{Ziemann2022}
I.~M. Ziemann, H.~Sandberg, and N.~Matni, ``Single trajectory nonparametric learning of nonlinear dynamics,'' in \emph{conference on Learning Theory}, 2022, pp. 3333--3364.

\bibitem{Cortes2016}
C.~Cortes, V.~Kuznetsov, M.~Mohri, and S.~Yang, ``Structured prediction theory based on factor graph complexity,'' \emph{Advances in Neural Information Processing Systems}, vol.~29, 2016.

\bibitem{Bishop}
C.~M. Bishop, \emph{Pattern Recognition and Machine Learning (Information Science and Statistics)}.\hskip 1em plus 0.5em minus 0.4em\relax Berlin, Heidelberg: Springer-Verlag, 2006.

\bibitem{Golowich2019}
N.~Golowich, A.~Rakhlin, and O.~Shamir, ``Size-independent sample complexity of neural networks,'' in \emph{Proceedings of the 31st Conference On Learning Theory}, vol.~75, 2018, pp. 297--299.

\bibitem{Stickler2016}
B.~Stickler and E.~Schachinger, \emph{Basics Concepts in Computational Physics}.\hskip 1em plus 0.5em minus 0.4em\relax Springer, 2016.

\bibitem{Gede2013}
G.~Gede, D.~L. Peterson, A.~S. Nanjangud, J.~K. Moore, and M.~Hubbard, ``Constrained multibody dynamics with python: From symbolic equation generation to publication,'' in \emph{International Design Engineering Technical Conferences and Computers and Information in Engineering Conference}, vol. 55973, 2013, p. V07BT10A051.

\bibitem{Gouk2021}
H.~Gouk, E.~Frank, B.~Pfahringer, and M.~J. Cree, ``Regularisation of neural networks by enforcing lipschitz continuity,'' \emph{Machine Learning}, vol. 110, pp. 393--416, 2021.

\end{thebibliography}

\end{document}